# The Jacobian as a measure of planar dose congruence

L. D.Paniak[1] and P. M. Charland[2]

[1]4π Solutions Inc, Kitchener, Ontario

[2]Grand River Hospital, Kitchener, Ontario and the University of Waterloo, Waterloo, Ontario

**Abstract**

We propose a new starting point for comparing dose distributions using a Jacobian-based measure. The measure is normalization independent, free of tunable parameters, bounded and converges to a unique value when comparing unrelated dose distributions. We present a preliminary demonstration of the sensitivity and general characteristics of this measure.

## 1. Introduction

The current state of the art in intensity-modulated radiation therapy (IMRT) for comparing planar dose distributions (Bakai *et al* (2003), Jiang *et al* (2006), and references therein, Low *et al* (1998), Moran *et al* (2005), and Van Dyk *et al* 1993) is to use measures involving either a Euclidean distance ($L_2$-sum of the squared differences) or the Manhattan distance ($L_1$-sum of the absolute differences). Research from Childress *et al* (2005) has pointed out that such scalar metrics do not have sufficient delivery error detection rates for IMRT QA. According to the same study, the visual inspection of the 2D planar doses was believed to have a nearly perfect error detection rate. Consequently, visual inspection of film was recommended in conjunction with the scalar metrics (Childress *et al*, 2005). Moreover, a recent survey of IMRT QA by Nelms and Simon (2007) has shown that standards for comparing planar dose distributions have yet to be defined. Typically the combined 3% and 3 mm criteria are used along with tools based on scalar metrics such as dose difference, distance-to-agreement and gamma index measures in informal and widely varied approaches to QA. Given these ambiguities, it may prove useful to explore new mathematical tools for objectively, quantitatively and automatically analyzing dose plans. For this purpose we propose a new starting point for comparing planar dose distributions based on the "cross-product", or Jacobian of the distributions.

Before considering new methods for comparing dose distributions that arise in circumstances such as IMRT QA, it is useful to collate a list of desirable features that quantitative, pair-wise measure of planar dose distributions should possess. In turn, these features can be classified in terms of how difficult they are to implement in practice. While this list is not exhaustive, we feel it gives a well-founded basis on which to develop a new paradigm for dose comparison.

Our aim is to provide a "measure(A,B)" of dose distributions A and B which has three relatively easily achievable qualities. First is normalization independence: measure(A,B) equals measure($\lambda$A, $\mu$B), where $\lambda$ and $\mu$ are constants. This feature will remove the need for normalizing dose distributions A and B. While absolute dose (normalization) is critical for a complete dose comparison, the proposed measure has the purpose of verifying isodose patterns. The measure (A,B) will avoid normalization bias by decoupling the comparison of (iso)dose patterns from global dose scaling. Such decoupling is otherwise not possible with current methods and it gives a new avenue for dose comparison. Second, the measure is bounded and measure(A,B) should have a natural limiting value when comparing images A and B which are completely unrelated. Thirdly, the measure has no tunable parameters, and requires no human input such as tolerances and search radius to bias its computation. A measure, which fulfills this last property, is uniquely suited for an automated analysis and may lead to better reproducibility.

More difficult to achieve, we also want non-locality when comparing the qualities of A and B dose distributions. Hence, the measure at a point contains information about the neighbourhood of the point.

Use of non-local data provides a richer set of objects to perform comparative operations on. Vector functions, such as the gradients of dose distributions, can be added or subtracted just as the (scalar) dose distribution can, but they also lend themselves to more complex operations such as the scalar or cross product. This provides a mathematical gateway for the comparison of patterns of isodoses between distributions A and B.

Even more difficult, we would like measure(A,B) to quantify some aspect of a visual analysis. In particular, it should cue on aspects of images noticed by human vision, which are typically geometric in nature. Finally, the most difficult goal to achieve is defining a pass or fail criteria. We acknowledge that an ultimate framework for planar dose analysis has to be clinically meaningful. This last desire is inherently subjective and beyond the scope of this paper, though we will have some suggestions for how to address it in our conclusions.

**2. Theory**

Given two scalar dose distributions *A* and *B*, defined on the same (x, y) plane, the tangent field of each dose distribution is given by a 2-vector. It is meaningful to compare tangent vectors $\vec{\nabla}A$ and $\vec{\nabla}B$ through a cross product,

$$J_{AB} = |\vec{\nabla}A \times \vec{\nabla}B| = \left(\frac{\partial A}{\partial x}\frac{\partial B}{\partial y} - \frac{\partial A}{\partial y}\frac{\partial B}{\partial x}\right) = |\vec{\nabla}A||\vec{\nabla}B|\sin\theta_{AB} \quad (1)$$

where $\theta_{AB}$ is the angle between vectors $\vec{\nabla}A$ and $\vec{\nabla}B$. The Jacobian $J_{AB}$ is a purely geometric measure. It is a point-wise measure of parallelism between tangents of isodoses in dose distributions A and B. It is convenient to divide the Jacobian by the norm of the dose gradients in order to produce a normalization-independent measure:

$$\sin\theta_{AB} = \frac{J_{AB}}{|\vec{\nabla}A||\vec{\nabla}B|} \quad (2)$$

The Jacobian index $j_0[A,B]$ is obtained from the integration of the normalized Jacobian measure over all points (x, y) and dividing by the overall area.

$$j_0[A,B] = \langle|\sin\theta_{AB}|\rangle = \frac{1}{area}\int_{x,y}|\sin\theta_{AB}|dxdy \quad (3)$$

Sometimes it is useful to consider the field θ(x,y) not as a dependent variable of the coordinates, but rather an independent one. With this re-interpretation, Equation 3 is equivalent to

$$j_0[A,B] = \int_{-\pi/2}^{\pi/2}|\sin(\theta)|\rho(\theta)d\theta \quad (4)$$

where ρ(θ) is a normalized probability density, periodic in θ: ρ(θ)= ρ(θ+π). The density ρ is essentially the Jacobian due to the change of variables from (x,y) to θ. This density gives a one-dimensional presentation of the distribution of angles between isodose lines in distributions A and B. It also provides a framework in which to examine two interesting cases:

1) When comparing two dose distributions that are completely unrelated, we expect that the angles between isodose lines will be broadly distributed. On average, this distribution is a uniform constant, $\rho_{random}(\theta)=1/\pi$ and the index $j_0[A,B]$ tends to a particular, calculable value:

$$j_0[A,B]^{rand} = \frac{1}{\pi} \int_{-\pi/2}^{\pi/2} |\sin\theta_{AB}| d\theta = \frac{2}{\pi} \approx 0.637 \qquad (5)$$

Numerical experiments comparing a representative IMRT composite dose distribution to a set of 25 random dose distributions generated by Fourier sums with random modes, frequencies and phases, gave a mean value for $j_0[A,B]$ of 0.612 with a standard deviation of 0.044.

2) If $\rho(\theta)$ is peaked at some large angle ($\theta \sim \pi/2$ radians), an appropriate rotation of one dose distribution will result in a lower value of $j_0[A,B]$. Conversely, a value of $j_0[A,B] > 2/\pi$ could be interpreted as a signal that a rotation can improve the correspondence of dose distributions. In a simple thought experiment, comparing two dose distributions, one with vertical isodose lines and a second with horizontal isodoses gives: $j_0[A,B]=1$. A rotation of one of the distributions by $\pi/2$ radians leads to a perfect match: $j_0[A,B_{rotated}]=0$.

Several generalizations of the index are possible. The first involves a "window function" which essentially calculates $j_0[A,B]$ in the vicinity of an isodose line of value $D_0$ :

$$j_0[A,B;D_0,\sigma_0] = \left\langle e^{-(D_A-D_0)^2/2\sigma_0^2} |\sin\theta_{AB}| \right\rangle \qquad (6)$$

In a similar fashion, the measure can be weighted by functions of dose gradients in order to isolate regions where dose changes quickly.

**3. Method**

We compare a reference IMRT planar dose distribution within Mathematica (Wolfram Research Inc., Champaign, Illinois) to various target scenarios calculated by the treatment planning software (Pinnacle, Philips Inc.) to mimic errors (Fig. 1). The planar dose in integer values from the treatment planning system is exported in ASCII format at 0.025 mm pixel resolution. Integrated measures of absolute dose difference $\delta[A,B]$, gamma index $\gamma[A,B]$, and the Jacobian index $j_0[A,B]$ are computed, all normalized to dose area under consideration. The gamma index is computed with 3% and 3 mm tolerances and a search radius of 6.5 mm. XV film measurements (Eastman Kodak Co., Rochester, NY) from irradiation on a 120 leaf MLCs Clinac-21EX (Varian Medical Systems, Inc., Palo Alto, CA) are added to the comparison. Films were pinpricked for registration purposes. Irradiated films were scanned with a VXR-16 Dosimetry Pro scanner (VIDAR systems Corporation, Herndon, VA) at 142 dpi and 16-bit depth and converted to dose.

The 16-bit depth resolution is required for accurate gradient calculation. To remove the noise in the film, we have used the band-pass filter option from ImageJ (Rasband 1997-2007). Structures larger than 2000 pixels (about twice the length of the image) were filtered as well as structures smaller than 15 pixels. This filtering was used to remove mild speckling on the original film data. This was necessary due to the fact that our algorithm requires the calculation of gradients. The results for the Jacobian measure were stable as the filtering was introduced. The gradient of the dose image is essentially an edge detection technique that is commonly used in imaging. Further details of the calculation are contained in a sample Mathematica code distributed at: http://www.fourpisolutions.com/projects/jacobian, along with dose distribution data used in this paper.

## 4. Results and Discussion

Calculation of a Jacobian index is completed in a few dozen seconds on a modern PC. The index increased commensurately with the magnitude of the deformation errors introduced in the target dose B (Table 1). The measures $\delta[A,B]$ and $\gamma[A,B]$ showed a better agreement with the 15 MV distribution scenario than with the film. Normalization bias and the point-wise nature of these measures, as well as the tunable parameters of the gamma index, are expected to impact these comparisons. The worst score (highest value) for the Jacobian index is in the collimator rotated 90° scenario, which also coincides with what visually appeared to be the worst matched patterns of isodoses. In figure 2, the Jacobian index $j_0[A,B]$ is shown to increase monotonically with the increase gantry angles and the resulting decrease in congruence between the reference and realized planar dose distributions.

## 5. Conclusions

The definition of procedures and criteria for judging planar dose distributions are ambiguous. This situation is due, in part, to the lack of a canonical measure for comparing dose distributions. Current measures, based on point-wise Manhattan or Euclidean distances, possess several shortcomings including normalization dependence, the need for user-defined input parameters and possibly the fundamental inability to discern fine differences between dose distributions. We propose another tool. This is only a preliminary step toward developing a comprehensive framework for making clinical judgment. With no tunable parameters and normalization independence, an immediate application of the Jacobian measure would be to use the index $j_0[A,B]$ in a preliminary, automated assessment tool for comparing dose distributions. Such a measure would allow for an unbiased initial assessment of isodose patterns. The consideration of dose magnitude (absolute dose), which is critical, is to be considered at a later stage of the assessment. The Jacobian measure can be weighed, overlaid or displayed as histogram and integration can be based on delineated regions of interests.

What is missing from our presentation of the Jacobian measure is a full framework for assessing dose distribution and a definitive "pass-fail" criterion. Naturally, such a definition of criterion involves discussion of whether a plan is acceptable or unacceptable in a clinical sense. These adjectives do not spontaneously lend themselves to quantification in a mathematical sense. Given this realization, we propose to define the pass-fail criteria for comparing dose distributions through a statistical process where the value of an index is calibrated against the best dose comparison mechanisms currently known, visual inspection. Implementation of such a calibration scheme would require polling physicists on the ranking of a standard set of isodose patterns. Such data would allow one to fine-tune (a set of) measures to provide an automated comparison system.

In summary, the proposed Jacobian measure is not an extension of current concepts used but rather a new perspective on the problem of comparing dose distribution. Future work is to explore the potential of the Jacobian measure and its combination with other tests in an effort to develop a complete framework for clinical decision.

**Email:**

ldpaniak@fourpisolutions.com

paule.charland@grhosp.on.ca

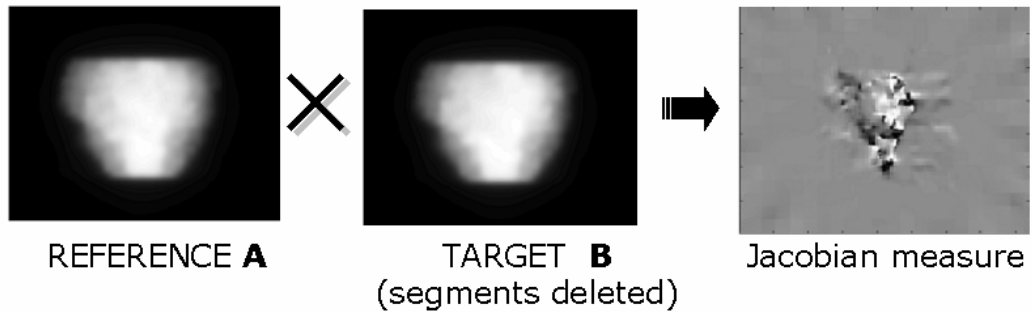

**Figure 1.** Illustration of the Jacobian measure $|sin\ \theta_{AB}|$ generated from two different dose distributions.

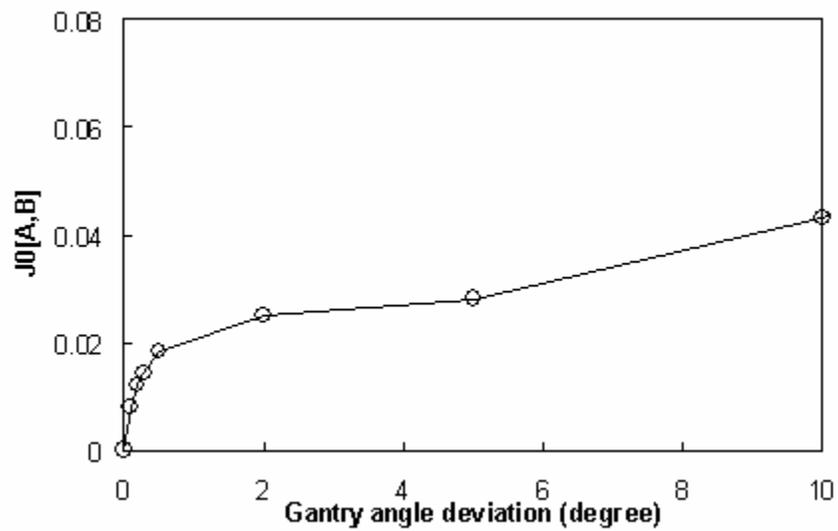

**Figure 2.** The Jacobian index of planar doses A and B increases monotonically with gantry angle deviation. Reference *A* is a dose distribution with gantry 0°. Planar target doses *B* are generated from the same beam as A except for different gantry angles.

**Table 1.** Comparison of several integrated, area normalized, measures: dose difference $\delta[A,B]$, Gamma index $\gamma[A,B]$ and $j_0[A,B]$. Reference planar dose distribution *A* (Fig.1) is calculated in Pinnacle for a 6MV photon beam at collimator and gantry angles of 0°. The first three scenarios of Target B are calculated in Pinnacle to simulate various errors. The fourth scenario corresponds to a film irradiated according to the configuration used to calculate reference dose A.

| Measure / Target B | $\delta[A,B]$ | $\gamma[A,B]$ | $j_0[A,B]$ |
|---|---|---|---|
| 1. Many MLC segments missing as compared to reference beam. | 3.5821 | 0.0108 | 0.1259 |
| 2. Beam identical to reference with the exception that 6 MV photons are replaced with 15MV. | 1.3441 | 0.0036 | 0.0918 |
| 3. Beam identical to reference with the exception that the collimator is rotated to 90°. | 15.7927 | 0.0469 | 0.4672 |
| 4. Film measurement of the reference beam. | 2.3425 | 0.0052 | 0.0762 |